\documentclass[comsoc,journal]{IEEEtran}
\usepackage{amsmath,amsfonts}
\usepackage{amsmath,amssymb,epsfig,xcolor}
\usepackage{algorithmic}
\usepackage{algorithm}
\usepackage{array}
\usepackage{textcomp}
\usepackage{stfloats}
\usepackage{url}
\usepackage{verbatim}
\usepackage{graphicx}
\usepackage{subfigure}
\usepackage[subfigure]{tocloft}
\usepackage{cite}
\usepackage{pifont}
\usepackage{tabularx,multirow,array}
\usepackage{booktabs}
\usepackage{diagbox}
\usepackage{makecell}
\usepackage{enumerate}
\usepackage{xcolor}
\usepackage{float}
\newcommand{\whiteding}[1]{\ding{\numexpr171+#1\relax}}

\begin{document}

\title{Trust-Worthy Semantic Communications for the Metaverse Relying on Federated Learning}

\author{Jianrui Chen,~\IEEEmembership{Student member,~IEEE,}
Jingjing~Wang,~\IEEEmembership{Senior Member,~IEEE,}
Chunxiao~Jiang,\\~\IEEEmembership{Senior Member,~IEEE,}
Yong~Ren,~\IEEEmembership{Senior Member,~IEEE,}
and Lajos~Hanzo,~\IEEEmembership{Life Fellow,~IEEE,}
        }

\markboth{IEEE WIRELESS COMMUNICATIONS}%
{Shell \MakeLowercase{\textit{et al.}}: A Sample Article Using IEEEtran.cls for IEEE Journals}

\maketitle

\newcommand\blfootnote[1]{%
\begingroup
\renewcommand\thefootnote{}\footnote{#1}%
\addtocounter{footnote}{-1}%
\endgroup
}

\blfootnote{This work of Jingjing Wang was partly supported by the Young Elite Scientist Sponsorship Program by the China Association for Science and Technology under Grant No. 2020QNRC001, and partly supported by the Fundamental Research Funds for the Central Universities. (\textit{Corresponding author: Jingjing Wang.})}
\blfootnote{J. Chen and J. Wang are with the School of Cyber Science and Technology, Beihang University, Beijing 100191, China (e-mail: chenjr2020@foxmail.com, drwangjj@buaa.edu.cn).} 
\blfootnote{C. Jiang is with Beijing National Research Center for Information Science and Technology (BNRist), Tsinghua University, Beijing, 100084, China (e-mail: jchx@tsinghua.edu.cn).}
\blfootnote{Y. Ren is with the Department of Electronic Engineering, Tsinghua University, Beijing, 100084, China (e-mail: reny@tsinghua.edu.cn).}
\blfootnote{L. Hanzo is with the School of Electronics and Computer Science, the University of Southampton, Southampton SO17 1BJ, U.K (e-mail: lh@ecs.soton.ac.uk).}

\begin{abstract}
As an evolving successor to the mobile Internet, the Metaverse creates the impression of an immersive environment, integrating the virtual as well as the real world. In contrast to the traditional mobile Internet based on servers, the Metaverse is constructed by billions of cooperating users by harnessing their smart edge devices having limited communication and computation resources. In this immersive environment an unprecedented amount of multi-modal data has to be processed. To circumvent this impending bottleneck, low-rate semantic communication might be harnessed in support of the Metaverse. But given that private multi-modal data is exchanged in the Metaverse, we have to guard against security breaches and privacy invasions. Hence we conceive a trust-worthy semantic communication system for the Metaverse based on a federated learning architecture by exploiting its distributed decision-making and privacy-preserving capability. We conclude by identifying a suite of promising research directions and open issues.
\end{abstract}


\section{Introduction}
\IEEEPARstart{S}{ince} the first appearance of the term `Metaverse' in Neal Stephenson's science fiction novel, its concept has been linked with new connotations, such as 3D virtual worlds, a second life, life-logging, etc. At the time of writing, the Metaverse may be viewed as an immersive spatio-temporal and self-sustaining shared virtual space capable of seamlessly blending the physical, human and digital worlds. Driven by recent advances in emerging technologies such as artificial intelligence (AI), the sixth-generation (6G) systems and blockchain, the Metaverse is stepping out of the world of fiction into reality. For instance, 1) virtual reality (VR) aims for making the virtual world more realistic via advanced computing technologies, exemplified by Google Earth, Microsoft Virtual Earth, 2) augmented reality (AR) aims for making the real world more virtual via advanced computer vision technologies, illustrated as Google Project Tango and Apple ARkit. During the COVID-19 era, it became more popular to run businesses, education and entertainment virtually. Hence, the Metaverse may be viewed as an evolving next-generation Internet paradigm.\par
However, the aforementioned applications tend to provide physical-virtual services for a single user without relating to other users or Internet of Things (IoT) devices. Hence, they are still far from realizing the full vision of an immersive, real-time, and seamlessly connected Metaverse. The bottleneck of implementing the Metaverse lies in flawlessly yet efficiently transmitting and processing an unprecedented amount of heterogeneous multi-modal and interference-contaminated data while supporting billions of users. To meet these stringent requirements, ultra-high throughput ($\geq$100GB/s), ultra-low latency ($<$0.1ms) and high reliability ($\geq$99.9999\%) must be maintained. In this context, Meng \textit{et al.}\cite{she2023} proposed a sampling, communication and prediction co-design framework for synchronizing the real-world devices and their digital models with high reliability.\par
Additionally, the personal data involved in developing the Metaverse must be protected. For instance, the high-end wearable VR/AR devices of the users may be endowed with motion sensing, artificial intelligence algorithms supporting different sensors in collecting, analyzing and conveying the users' facial expression variations, body movements, speech prosody as well as their surrounding environment. To accurately reflect these nuances on the virtual avatar, these devices and AI models will memorize some personal features and private data. Furthermore, interacting with other users requires that each user has a unique identity, hence they might be tracked, leading to erosion of privacy. To address this aspect, Wang \textit{et al.}\cite{wang2023} proposed a novel distributed metaverse architecture and presented an in-depth survey of security and privacy preservation measures conceived for the distributed metaverse architecture considered. And Lin \textit{et al.}\cite{blockchain} proposed a unified blockchain-semantic ecosystem framework to provide security services.\par
In a nutshell, there are a pair of salient issues to be addressed: 
\begin{itemize}
  \item \textbf{Communication capacity and efficiency:} Ultra-massive access and real-time synchronization impose more stringent requirements on capacity and efficiency than that of 5G. In this context, semantic communication is gradually becoming the scheme of choice to address these issues. In contrast to the traditional Shannonian paradigm, semantic communication extracts the most salient information features and only transmits the information that is the most relevant to the specific tasks at the receiver. This results in significant reduction in data traffic. 
  \item \textbf{Privacy preservation:} In the Metaverse, the various wearable devices collect almost all clients' information and transmit these data to the cloud or servers, which might compromise the clients' general privacy, exposing their identity, location, etc. However, semantic communication is not a privacy-preserving scheme, where distributed users share a massive amount of data. Therefore, it is an extra challenge to make semantic communications trust-worthy. 
\end{itemize}

To solve the above problems, we conceive a novel semantic communication architecture for the Metaverse, with particular focus on the communication and privacy-preservation issues. Explicitly, we propose a privacy-preserving multi-user semantic communication system based on federated learning (FL) to address the associated privacy leakage. For the former problem, on one hand adopting semantic communications to only transmit task-oriented information reduces the communication overheads eminently on the one hand. On the other hand, the proposed multi-user semantic communication system will harness both intelligent radio (IR) solutions and state-of-the-art multiple access techniques for further enhancing the communication capacity. For the latter problem, by leveraging differential privacy (DP) and knowledge distillation (KD), our architecture has become privacy-preserving and robust, as we will demonstrate by our simulation results.\par
The rest of the article is outlined as follows. Section \uppercase\expandafter{\romannumeral2} provides an overview of semantic communications and highlights the privacy threats in the Metaverse. Then a detailed FL-aided semantic communication architecture will be given, followed by a range of open issues on privacy-preserving semantic communications, before we conclude.

\section{Semantic Communications for the Metaverse}

\begin{figure*}[!t]
  \centering
  \includegraphics[width=0.9\textwidth]{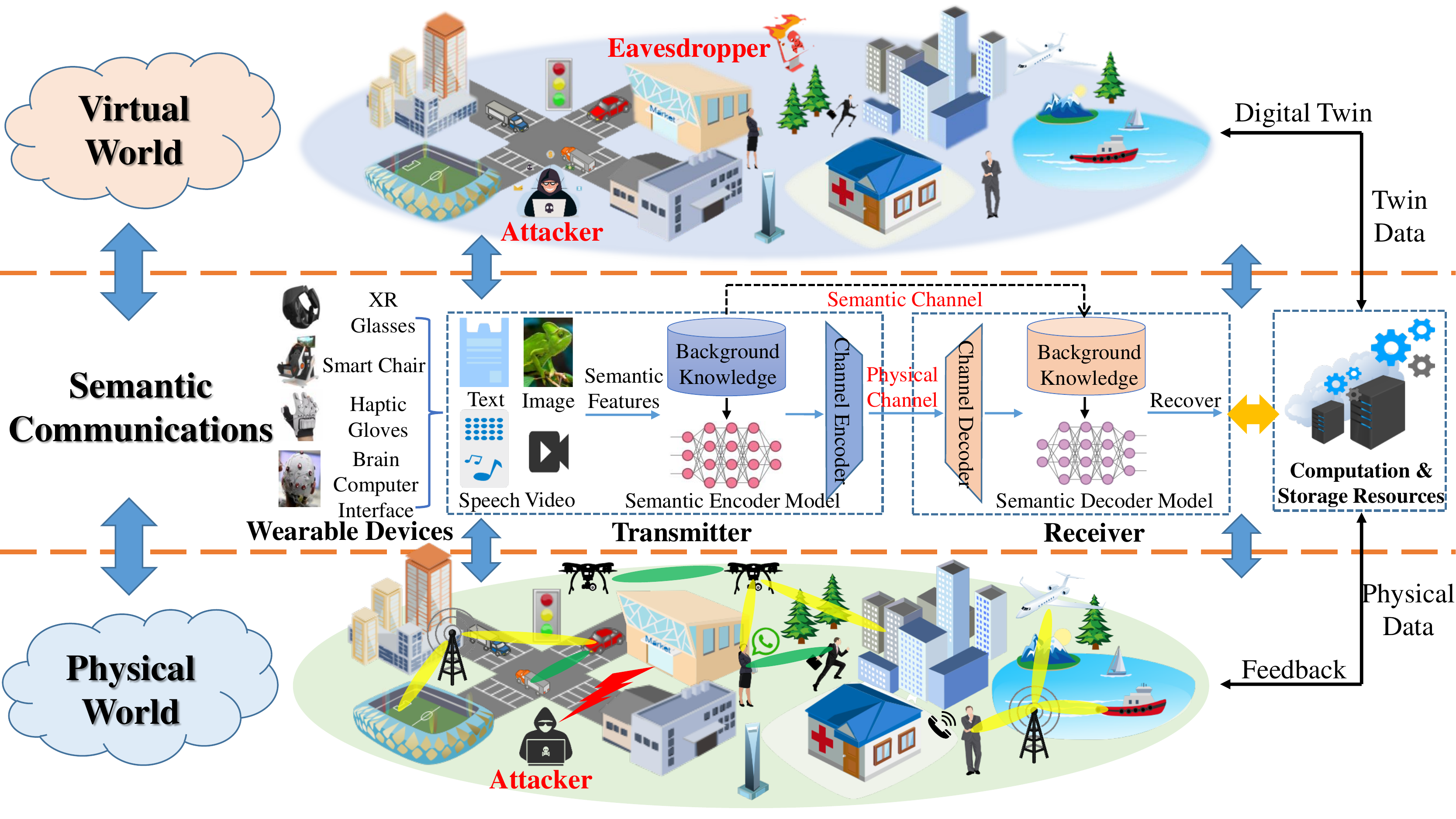}
  \caption{The architecture of the Metaverse assisted by semantic communications}
  \label{system_model}
\end{figure*}
In contrast to Shannon's classical communication paradigm of avoiding any loss of information, the goal of semantic communication is only to reconstruct the indispensable information, which the completion of a specific task requires instead of perfectly restoring the original data collected from the source at the transmitter\cite{semantic1}. On this basis, the semantic transceivers have matched background knowledge bases (KBs) and only exchange task-oriented information extracted from the raw data. Hence the communication traffic and bandwidth requirements are significantly reduced." Based on the characteristics of semantic communications, Fig. \ref{system_model} depicts a blueprint of the Metaverse assisted by semantic communications.\par
As shown in Fig. \ref{system_model}, through wearable devices, such as XR glasses, smart chairs, haptic gloves and brain computer interfaces, people can construct their avatars and reflect them in the virtual world. Semantic communication, as a broad bridge, connects the physical and the virtual world of billions of users, which creates an unprecedented data Tsunami. In this section, the general principles of semantic communication will be discussed. 
\subsection{\textbf{End-to-End Semantic Communication Framework}}

\begin{figure*}[!t]
  \centering
  \includegraphics[width=0.8\textwidth]{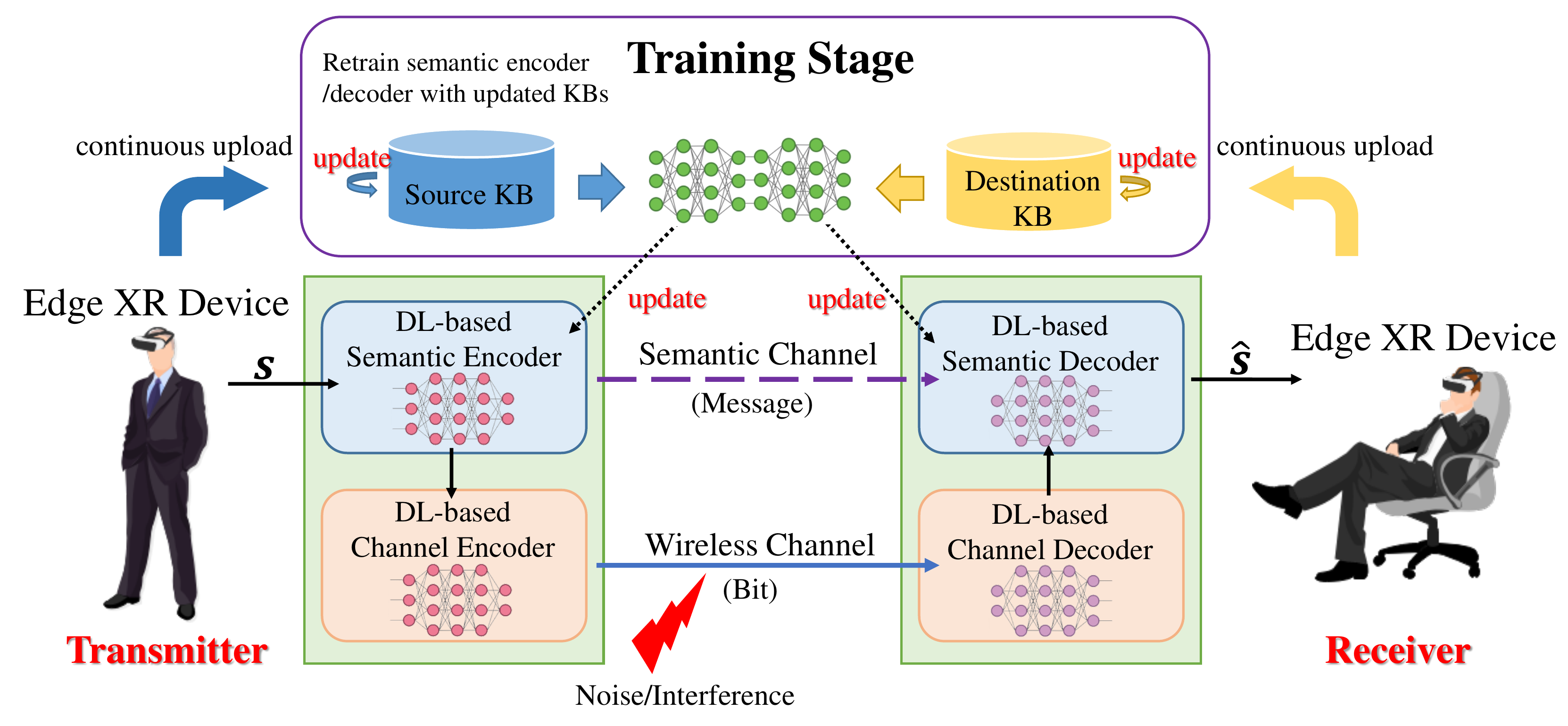}
  \caption{An illustration of an end-to-end semantic communication system.}
  \label{e2e_system}
\end{figure*}

With the development of deep learning (DL), an increasing number of DL-based semantic communication systems have been proposed, which can automatically learn to extract and transmit semantic information. For instance, Xie \textit{et al.} \cite{SCxie} proposed a semantic communication system based on deep transfer learning for text transmission, while Wang \textit{et al.} \cite{attention} conceived a reinforcement learning based textual data transmission mechanism. Furthermore, Kalfa \textit{et al.} \cite{kalfa2021towards} designed a semantic signal processing framework, which can be flexibly changed between specific tasks. Observe in Fig. \ref{e2e_system} that by leveraging DL, we construct a general end-to-end (E2E) semantic communication system, where both the transmitter and receiver can be humans, machines or other intelligent mobile devices. The transmitter should have the ability to extract the semantic features of the source messages and encode these features into symbols (bits) for transmission. The receiver should be able to interpret and recover the messages sent by the transmitter. Next, we will give an overview of the main components of an E2E semantic communication system.
\begin{itemize}
  \item \textbf{Background knowledge base:} Extracting sufficient features or `meaningful' information from the source messages requires transceivers capable of constructing their own task-oriented background KBs. According to the different mobile users or tasks, the KBs can be classified into different clusters. Consider machine translation as an example. People who speak the same language (may English or Chinese, etc.) should be put into the same cluster for enhancing the translation performance by analyzing their personal characteristics. Establishing self-contained KBs is a long-term process. Since the knowledge matching between the source and destination KBs is one of the most crucial factors, which is strongly related to the accuracy of the associated semantic interpretation, the transmitter and receiver have to share their KBs in advance. After sharing their KBs, the transceivers can successfully understand, compress and recover the information. However, sharing the KBs also has the risk of privacy leakage.
  \item \textbf{Semantic encoder and decoder:} In contrast to compressing information by source encoding and channel encoding to acquire an excellent performance in traditional wireless communication systems, a semantic encoder should extract the semantic features of the data. On the one hand, the semantic encoder should retain the accuracy in representing observed world as much as possible. On the other hand, it should also filter out the useless, irrelevant, and unessential information to minimize the data traffic. At the receiver, the channel decoder first decodes the received signal and then recovers the original messages via the semantic decoder by extracting the semantic features. These factors impose critical challenges on the semantic encoder and decoder, motivating us to design intelligent communication systems by considering the semantic meaning conveyed by the bits to enhance the accuracy and efficiency of communications. In this context, DL has shown great potential in the semantic representation of natural language \cite{NLP}, text transmission\cite{text}, speech transmission \cite{speech} and so on. 
  \item \textbf{Semantic and physical channels:} As shown in Fig. \ref{e2e_system}, there are two types of channels in semantic communication systems. The first type of channels are the wireless channels, which may suffer from noise, fading, and inter-symbol interference. In the past, researchers have invested substantial efforts into combating the physical channel impairments. The second type of channels are the semantic channels, which impose potential misunderstanding or interpretation errors, even in the absence of transmission errors.
\end{itemize}\par
However, although the aforementioned system is eminently suitable for an end-to-end communication, it will encounter numerous challenges in the Metaverse. Firstly, building perfect KBs is a long-term process and frequently updating the KBs will result in the trained semantic encoder and decoder to become outdated. Secondly, semantic communication is essentially a receiver-dominated communication form. Specifically, the transmitter's action of training a semantic encoder critically hinges on the specific type of information features the receiver needs. This means that the transceivers should make prior agreements and formulate rules based on the receivers' tasks, which is particularly challenging for the distributed the Metaverse users. Thirdly, the transmitted data is usually non-independent and identically distributed (non-i.i.d), which makes training the model hard. Finally, training the DL-based semantic encoder/decoder requires them to share their KBs, which may lead to privacy leakage. These challenges must be circumvented in distributed multi-user semantic communications in the Metaverse.

\subsection{\textbf{Privacy-Preserving Multi-user Semantic Communications}}
In the previous subsection, we have introduced several semantic communication concepts, where DL plays a critical role in feature extraction and communication. While much of the research into end-to-end semantic communications has focused on optimizing the local semantic encoder/decoder DL model, an equally important yet under-explored problem is the design of distributed multi-user semantic communications in the Metaverse. First of all, the connectivity density of 6G networks may escalate to $10^{8}$ devices per $km^{2}$, which requires a significant spectral-efficiency improvement. Non-orthogonal multiple access (NOMA) might be harnessed for improving the spectral-efficiency in semantic communication systems. More explicitly, the multi-user signals can be transmitted using the same frequency and time-slot via multiple antennas by exploiting the compelling spatial division multiple access (SDMA) philosophy separating the users based on their unique antenna/user-specific impulse response. However, the employment of accurate iterative channel estimation and data detection is vital for the reliable separation of users. As an explicit benefit of this iterative SDMA channel estimation and data detection SDMA the number of users supported may even be twice higher than the number of antennas. Furthermore, the source signal is usually non-i.i.d and may not be collected in the KBs. This will increase the difficulty of training the DL model for the encoder, but perfectly matched and complete KBs constitute a critical premises for training accurate semantic encoders/decoders. This means that the users connected to the Metaverse should share all their data and personal characteristics without reservation. This will expose users to eavesdroppers or other malicious agents. In fact, most users would prefer to protect their privacy at the expense of some communication performance erosion.\par

\begin{figure*}[!t]
  \centering
  \includegraphics[width=0.9\linewidth]{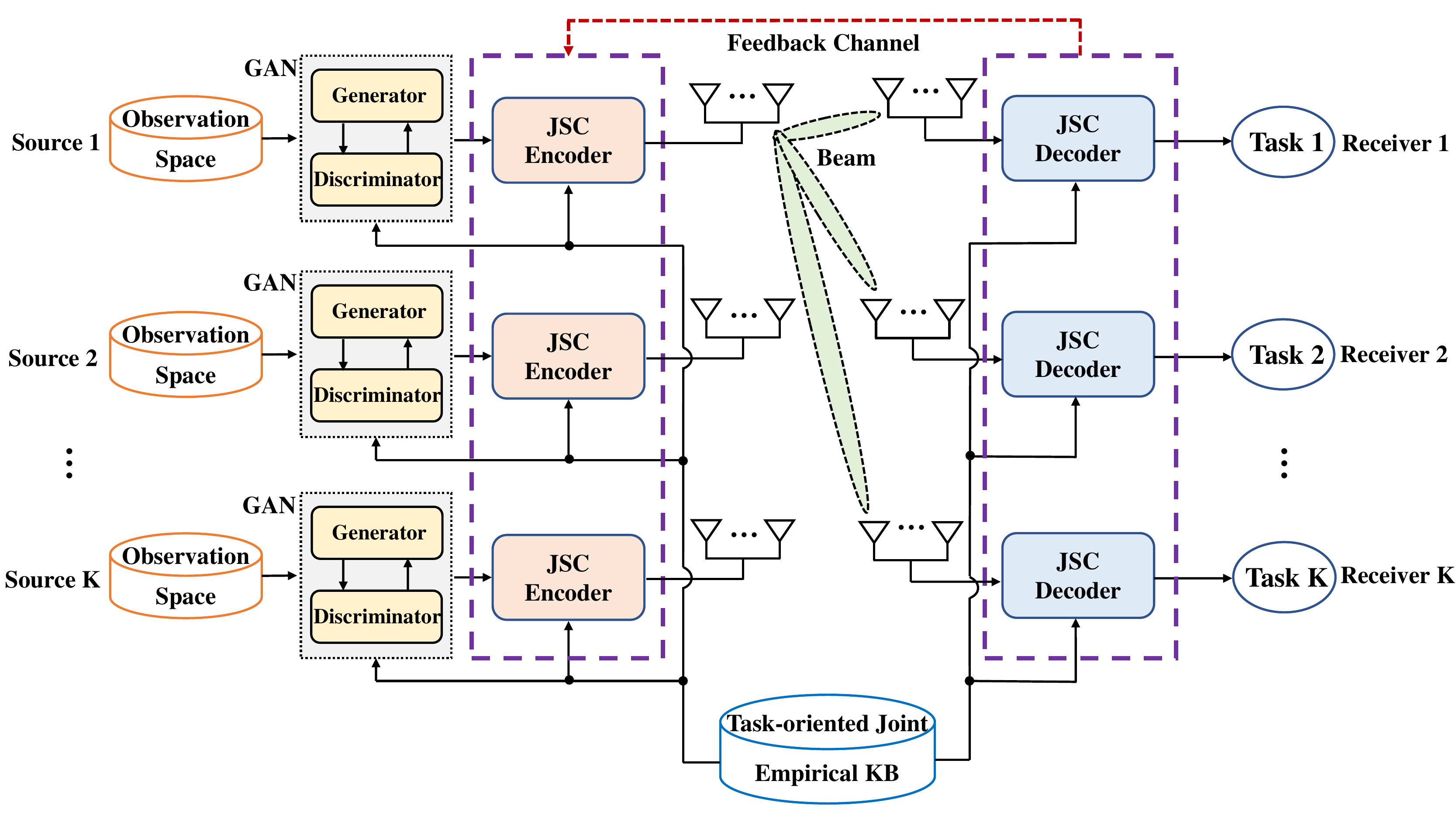}
  \caption{An illustration of a multi-user semantic communication system based on IR and GAN.}
  \label{mSC_system}
\end{figure*}

To solve the above challenges, we propose a secure multi-user semantic communication system based on the IR and generative adversarial network (GAN) shown in Fig. \ref{mSC_system}. It is evolved from the E2E system for overcoming the shortcomings mentioned above. In this system, we combine the semantic and channel encoders (or decoders) into joint semantic-channel (JSC) encoders (or decoders). By harnessing a deep neural network (DNN), an IR receiver in IR can estimate each client's channel state information (CSI) and separate multi-user signals. Furthermore, sophisticated transmitter and receiver beamforming relying on multiple-input multiple-output (MIMO), solutions can be used for improving the performance of multi-user semantic communication systems. These physical technologies are used to help build a seamlessly connected virtual world, where every device, avatar or user can interact with any other without any transmission medium. Moreover, we adopt GANs to transfer the recently observed data to the knowledge base without updating the KB and retraining the semantic coding networks via the domain adaptation (DA) technique of \cite{zhang2022deep}. This unique characteristic of GAN is not possessed by other neural networks. Specifically, we firstly build a task-oriented joint empirical KB based on the universal dataset shared by the transceivers. Although the joint KB is imperfect, refraining from uploading all the data protects the clients' personal privacy. Then by harnessing the DA network, the actually observed data will be pre-processed in GAN with the whole semantic coding models unchanged. As shown in Fig. \ref{mSC_system}, there is a generator and a discriminator in GAN. The generator is designed for transforming the observation data into a specific representation matched to the joint KB, while the discriminator is designed for distinguishing the data transformed by the generator and the data in the KB. The training of GAN will be continued until the discriminator succeeds in distinguishing the two kinds of data. Hence the DA neural network can be trained locally at the transmitter without communicating with the receiver. Above all, there are three principle advantages of this framework: 1) the non-i.i.d data can be converted into a similar representation of the joint KB to enhance the robustness of the semantic coding network, 2) the well-trained semantic models can be re-used, and 3) local privacy can be protected without being uploaded to the joint KB.  

\subsection{\textbf{Privacy Threats in the Semantic Communications Aided Metaverse}}
While using semantic communications in the Metaverse, user privacy, including both the local KB and the semantic coding parameters which contain personal data, behavior habits and other characteristics, is vulnerable to so-called Sybil attacks and distributed denial-of-service (DDoS) attacks, potentially resulting in semantic coding paralysis. Although the proposed architecture is capable of protecting the users' privacy by sharing data with reservation, it is far away from trust-worthy in the face of ubiquitous privacy threats. Privacy leakage may occur at each stage of semantic communications. In this subsection, we will highlight the privacy threats in the semantic communications assisted the Metaverse.\par
\textbf{1) Privacy leakage in data collection:} To seamlessly integrate the virtual world and the real world, the Metaverse will collect user profiling activities at a high resolution using advanced wearable devices, including eye movements, facial expressions, speech characteristics, various biometric features and even brain wave patterns. Then this heterogeneous data will be stored in different KBs. If the edge devices are attacked by malicious users, the sensitive data in the storage will be leaked\cite{table1}. In semantic communications, the privacy leakage during data collection occurs at the KBs, in form of IP spoofing and software hijacking.\par
\textbf{2) Privacy leakage in data transmission:} In semantic communication systems, the identifiable personal information collected from wearable devices will be compressed and transmitted via wireless channels. Similarly to traditional communications, most privacy attacks usually occur in this stage, but the information is usually encrypted and transmitted confidentially using secure multi-party computation. Nevertheless, attackers may still be able to decipher the messages. Owing to the fact that attackers have no prior knowledge about the KBs, semantic communication systems can help to protect the users' privacy, even if the messages have been eavesdropped upon. However, by leveraging DL, the adversaries have advanced methods to infer users' private data by differential and inference attacks\cite{table2}.\par
\textbf{3) Privacy leakage in data processing:} Given that semantic coding/encoding schemes constitute the most essential components in DL-based semantic communication, the parameters of the well-trained models determine how to compress the raw data at the encoder or recover the data at the decoder. Once the adversaries have access to the model parameters, they can infer the encoding/decoding model to recover the raw information via reconstruction attacks. Most research efforts concentrate on designing encryption schemes for the model parameters, but these methods are usually confidential\cite{table3}.\par 
A summary of privacy threats and countermeasures in semantic communications assisted Metaverse is presented in Table \ref{summary}. 
\begin{table*}
  \caption{{\color{black}{Summary of privacy threats/countermeasures of semantic communications in the Metaverse}}}
  \label{summary}
  \begin{center}
  \resizebox{\textwidth}{!}{
\begin{tabular}{c c c c c c}
  \toprule
  \multirow{3}{*}{\textbf{\normalsize Attack Stage}} & \multirow{3}{*}{\textbf{\normalsize Ref.}} & \multirow{3}{*}{\textbf{\normalsize Target Component}} & \multirow{3}{*}{\textbf{\normalsize Privacy Threat}} & \normalsize $\ast$ \textbf{\normalsize Purpose} & \multirow{3}{*}{\textbf{\normalsize Other Countermeasures}}\\
   & & & & \normalsize $\star$ \textbf{\normalsize Advantages} & \\
   & & & & \normalsize $\circ$ \textbf{\normalsize Limitation} & \\
  \midrule

  \multirow{3}{*}{\normalsize Data Collection} & \multirow{3}{*}{\normalsize [12]} & \multirow{3}{*}{\normalsize KBs, Edge Devices} & \multirow{3}{*}{\makecell[c]{\normalsize Hijack of wearable devices\\\normalsize Location tracking\\\normalsize User identification exposure}} & \makecell[l]{\normalsize $\ast$ A cloud-based user authentication scheme}  & \multirow{3}{*}{\makecell[c]{\normalsize  Access control\\\normalsize K-anonymity\\\normalsize L-diversity}}\\
  & & & & \makecell[l]{\normalsize $\star$ High resilience against wearable sensor node capture attacks } & \\
  & & & & \makecell[l]{\normalsize $\circ$ Highly complex and challenging to deploy in practical scenarios} & \\
  \midrule

  \multirow{3}{*}{\normalsize Data Transmission} & \multirow{3}{*}{\normalsize [13]} & \multirow{3}{*}{\normalsize Wireless Channels} & \multirow{3}{*}{\makecell[c]{\normalsize Linkage attack\\\normalsize Sybil Attack\\ \normalsize DDos attack}} & \makecell[l]{\normalsize $\ast$ A lightweight semantic privacy-preservation framework} & \multirow{3}{*}{\makecell[c]{\normalsize Secure multi-party computation\\\normalsize Blockchain\\\normalsize Homomorphic encryption}}\\
  & & & & \makecell[l]{\normalsize $\star$ Maintain privacy with high utility efficiency} & \\
  & & & & \makecell[l]{\normalsize $\circ$ Cannot ensure strong privacy} & \\
  \midrule

  \multirow{3}{*}{\normalsize Data Processing} & \multirow{3}{*}{\normalsize [14]} & \multirow{3}{*}{\normalsize Encoding/Decoding Models} & \multirow{3}{*}{\makecell[c]{\normalsize Byzantine attack\\\normalsize Reconstruction attack\\\normalsize Membership inference attack}} & \makecell[l]{\normalsize $\ast$ An adversarial encryption training scheme} & \multirow{3}{*}{\makecell[c]{\normalsize Tansfer learning\\\normalsize Local DP\\\normalsize Edge computing}}\\
  & & & & \makecell[l]{\normalsize $\star$ Guarantee the accuracy and prevent eavesdropping} & \\
  & & & & \makecell[l]{\normalsize $\circ$ High confidentiality but poor generality} & \\
  \bottomrule
\end{tabular}}
\end{center}
\end{table*}

\section{FL-aided Multi-user Semantic Communications}
As an emerging technique of jointly training the model for multiple participants in the field of secure machine learning, the popular distributed architecture of FL does not need to share any local data directly. It can make up for the lack of DL and extend the end-to-end semantic encoder/decoder model training to multi-user model training. By transmitting model parameters instead of the direct transmission of original data, its privacy-preserving capability has led to diverse applications. Above all, the unique features of FL can make real-time semantic communication possible by reducing the time delay caused by transmitting the original data. They also provide a more secure information interaction among multiple users.\par
However, FL still fails to provide sufficiently reliable privacy protection. In contrast to traditional centralized ML-aided privacy protection, the privacy attacks of FL are typically initiated by internal participants and recent studies have shown that the adversaries can infer the users' initial data through the model gradient information. For example, the members of a simple FL-aided semantic communication system can launch attacks by directly obtaining embedded representations of the shared KBs, gradients and other model parameters during the training of semantic encoders/decoders. This may also affect the training of models by replacing samples, changing gradients and even modifying loss functions. In this way, they can entice legitimate terminals into exposing more confidential information and launch both inference and reconstruction attacks. FL needs more participants for collaborative training and model sharing, but it lacks the corresponding identity confirmation mechanism and integrity guarantee, which makes it difficult to prevent `internal' leakage through membership inference attacks.\par
\begin{figure*}[!t]
  \centering
  \includegraphics[width=0.8\textwidth]{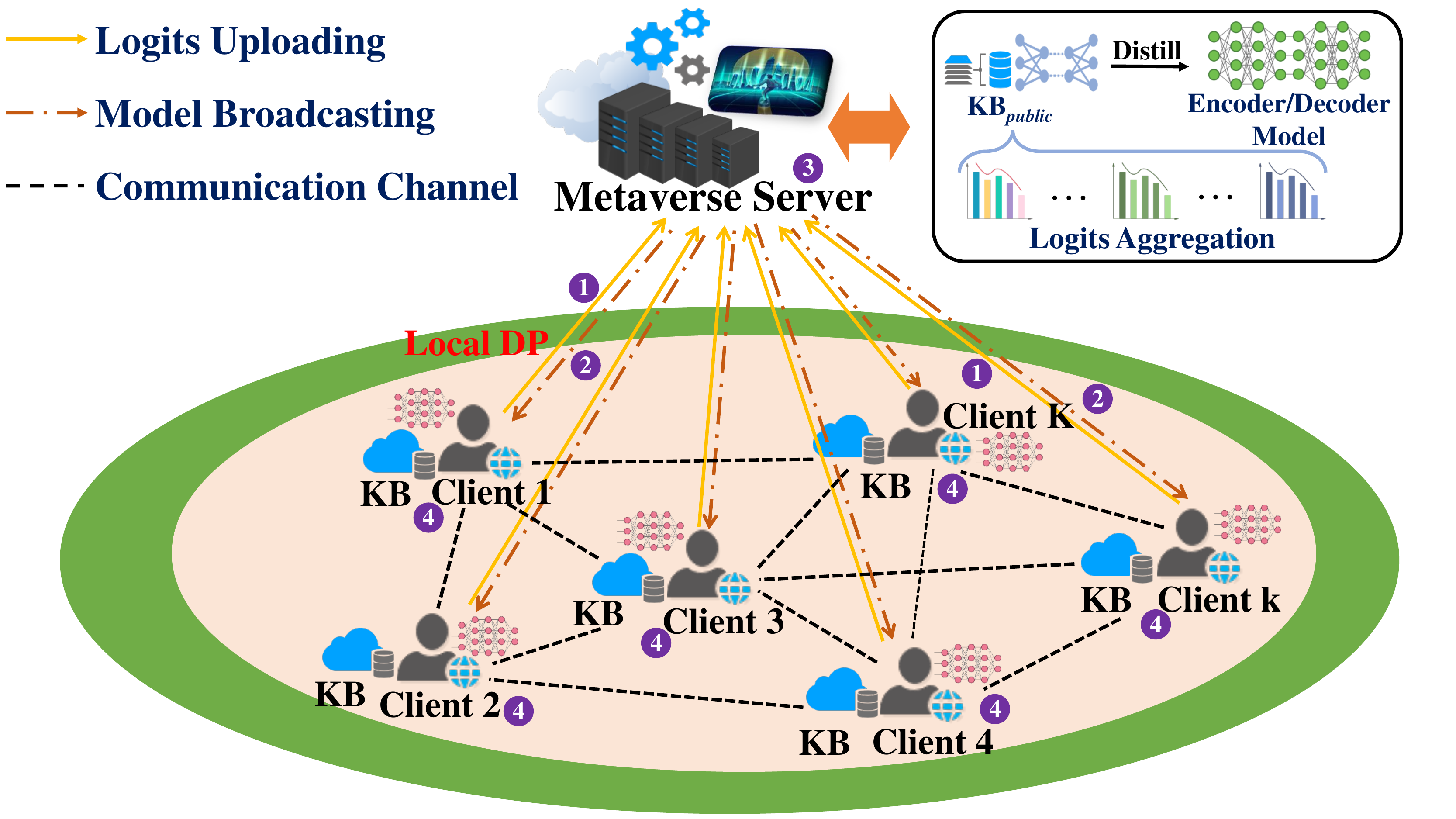}
  \caption{The architecture of semantic communications based on FL, which includes the following four steps: \whiteding{1}the client uploads local logits of its semantic encoder/decoder model; \whiteding{2}the Metaverse server broadcasts the distilled encoder/decoder model; \whiteding{3}the Metaverse server aggregates the uploaded logits from clients; \whiteding{4}the client updates its semantic encoder/decoder in accordance with the distilled parameters.}
  \label{FLSC}
\end{figure*}

 
In order to improve the privacy of FL, the privacy protection technologies currently used include secure multi-party computation, homomorphic encryption (HE), and DP. However, secure multi-party computation and HE have excessive computational and communication overheads as a price for their excellent privacy-preserving performance. Futhermore, DP is widely used in FL due to its rigorous mathematical foundations and efficient quantitative privacy analysis, which makes the FL architecture reliable. Hence, based on the multi-user semantic communication system of Fig. \ref{mSC_system}, we now propose the enhanced FL-aided multi-user semantic communication architecture of Fig. \ref{FLSC}.\par

As shown in Fig. \ref{FLSC}, we consider a scenario where several Metaverse clients interact with each other by semantic communication. Each client has his/her own different KBs according to the specific task. In the proposed architecture, the Metaverse server is assumed to be a trusted third party and the users will upload and store their public KBs in it. Note that the Metaverse server has built various, huge task-oriented joint empirical KBs, which are universal and accessible for each client. Based on these KBs, the users can obtain the encoder/decoder models and avoid the sharing of their KBs as well as local observation data. Similarly to the architecture designed in Fig. \ref{e2e_system}, the clients will periodically compute the local loss function of the encoder/decoder model relying on their real-time private observation-based datasets and the updated model parameters. In the traditional FL architecture, the model parameters will be uploaded and averaged, and then the Metaverse server will broadcast the aggregated model to the clients. However, the most widely-used federated average (FedAvg) algorithm may be inapplicable to FL-aided multi-user semantic communication systems, since it requires that the participants' encoder/decoder models should have the same neural network architecture. To update the heterogeneous models and reduce the communication overheads, KD is introduced as a novel aggregation method in our FL-aided multi-user semantic communication system\cite{distill} for distilling knowledge from a large and well-trained teacher model into a small student model. There is no need for KD to train the model from scratch, which substantially improves the training efficiency. Firstly, each client will compute the local encoder/decoder model output (logits) relying on the local private observation-based dataset and the public dataset in the server, followed by uploading them to the Metaverse server. Next, the Metaverse server will aggregate these logits and the aggregated logits will serve as the teacher encoder/decoder model for distilling a small and practical encoder/decoder model. Finally, the Metaverse server will broadcast the distilled encoder/decoder model to the clients via wireless channels. By harnessing federated distillation (FD), the FL architecture becomes more robust to the non-i.i.d nature of the dataset and to heterogeneous encoder/decoder model architectures.\par
\begin{figure}[!t]
  \centering 
  \subfigure[Accuracy versus different aggregation methods.]{
  \includegraphics[width=0.47\textwidth]{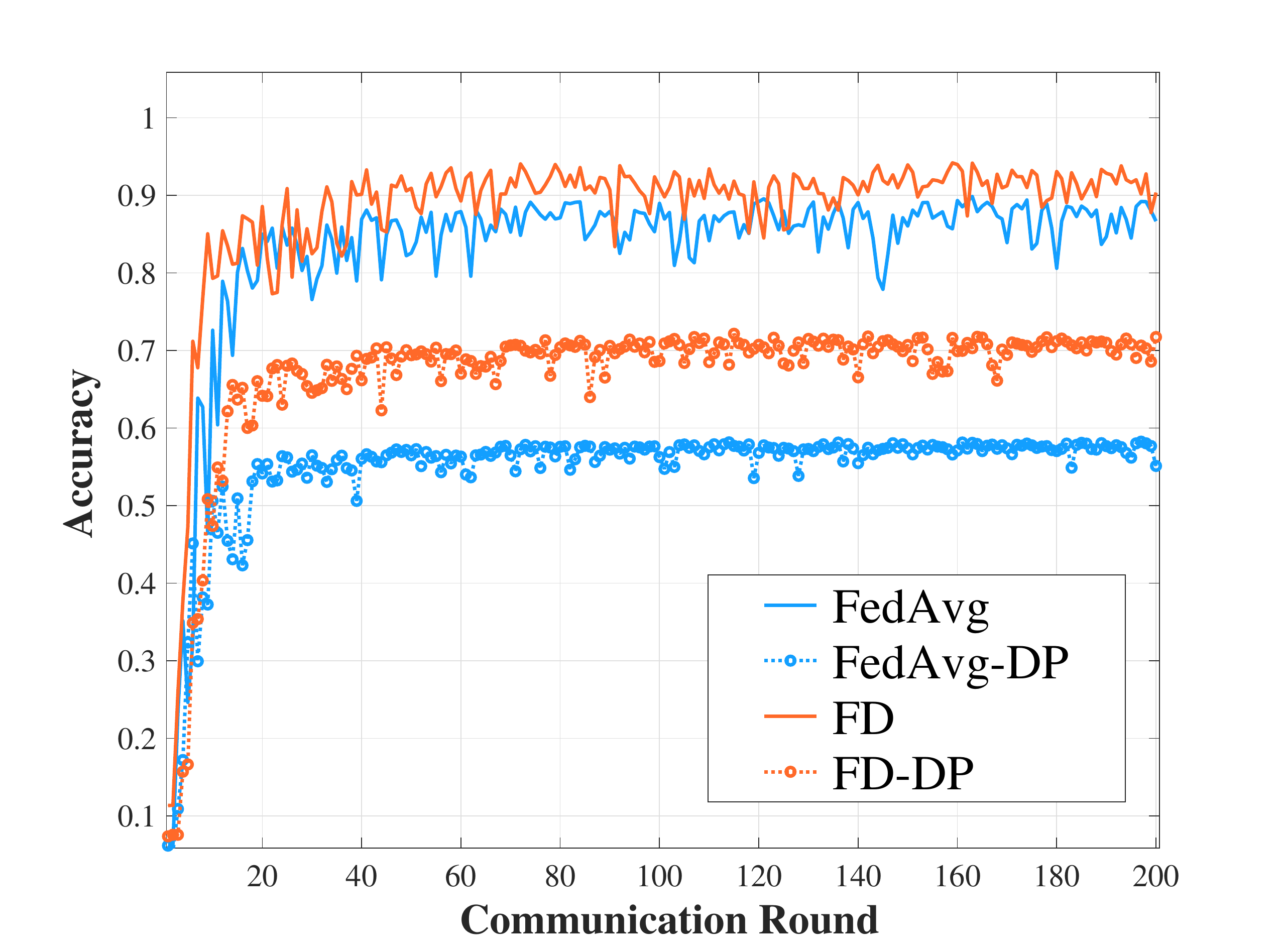}
  }
  \subfigure[Accuracy versus different privacy budget $\varepsilon$.]{
  \includegraphics[width=0.46\textwidth]{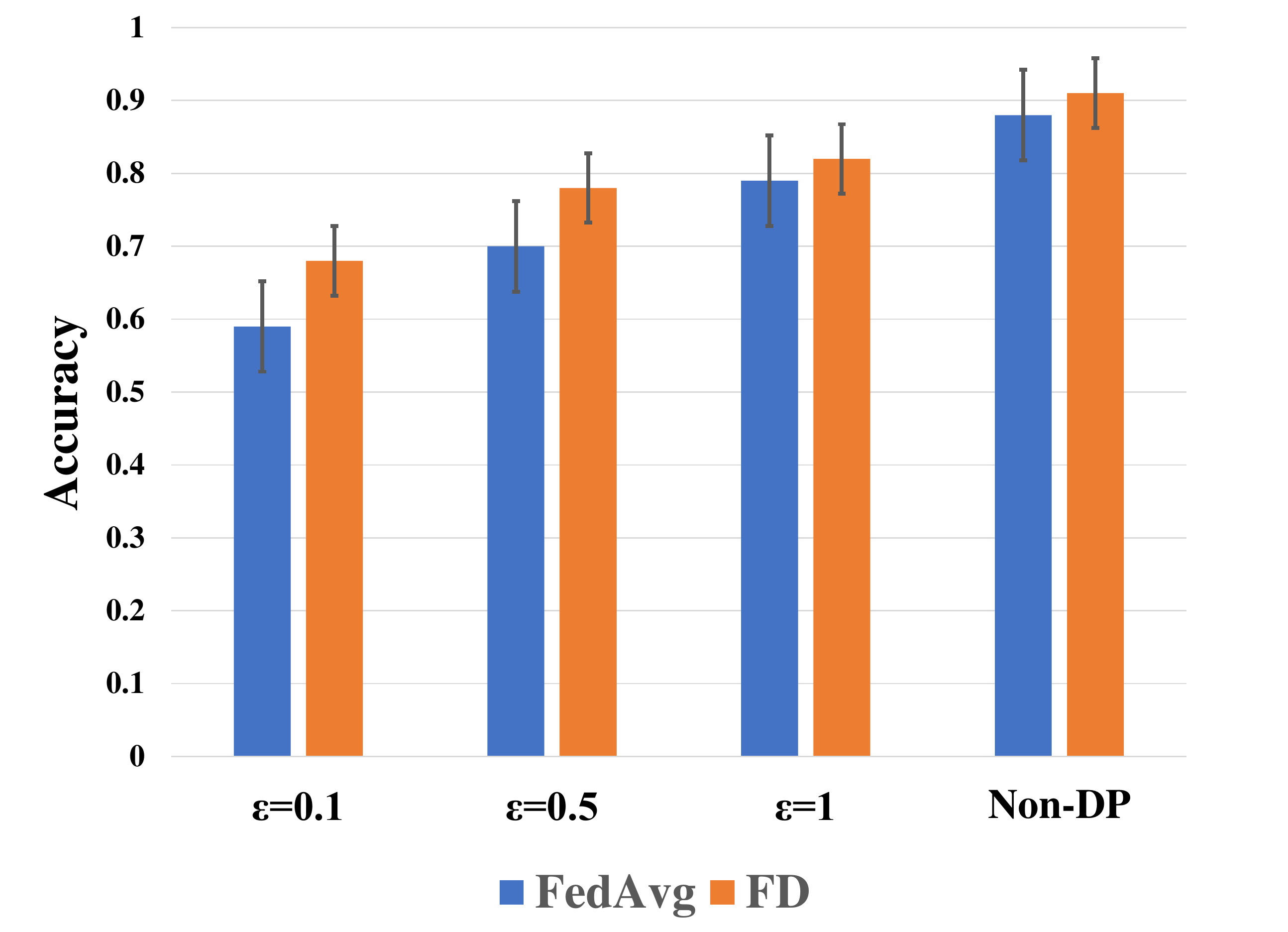}
  }%
  \caption{The simulation results under different FL architectures.}
  \label{simulations}
\end{figure}

Moreover, in order to avoid increasing the communication and computation cost, the clients opt for DP mechanisms to further enhance their local privacy before uploading the logits to the Metaverse server. Specifically, according to so-called $(\varepsilon-\delta)$-DP requirement, random noise is added to the uploaded logits for ensuring that the attacker cannot distinguish whether a certain client exists or uploads through known information. With the introduction of KD and DP, there will inevitably be a trade-off among the communication cost, privacy and accuracy. To demonstrate the efficiency and privacy-preserving performance of our architecture, we conduct experiments under a non-i.i.d MNIST dataset across 100 clients. Specifically, we partition the arranged MNIST subset of 30,000 samples into 100 groups of data slices with a size of 300, and then assign one slice to each client as their local private training data. The distillation dataset has on the other of 30,000 samples of the MNIST dataset. The neural network model is the popular CNN. In each communication round, 10 clients are selected randomly for participating in the learning process. The simulation results of Fig. \ref{simulations}(a) show that our architecture has a better accuracy than FedAvg under heterogeneous datasets. Although the accuracy is eroded, the privacy protection is improved by adopting DP. Moreover, observe in Fig. \ref{simulations}(b) that both methods' accuracy will be reduced upon reducing the privacy budget $\varepsilon$ and that our proposed architecture exhibits increased robustness. Above all, our proposed architecture is capable of maintaining high accuracy, while protecting privacy.

\section{Open Issues}
In this section, we list several major open issues for future investigations on how to construct a more trust worthy semantic communication architecture for the Metaverse.

\subsection{\textbf{Theoretical Analysis of Privacy-Preservation}}
The ultimate performance limits of the existing privacy-preserving mechanisms are unknown at the time of writing. Most of the literature adopts DP for the quantitative analysis of the privacy budget, which is typically carried out under strong simplifying assumptions concerning the malicious agents. But the definition of `privacy budget' is only applicable to DP.

\subsection{\textbf{Multi-Component Optimization of Semantic Communication}}
Here we advocate a novel FL-aided semantic communication architecture, with a particular focus on privacy protection. Harnessing GAN, KD and DP is capable of improving the privacy at the cost of accuracy erosion. Other existing mechanisms, such as multi-party computation and HE, will result in excessive communication and computation overheads. Therefore, low-cost and lightweight privacy-preserving strategies should be designed. Furthermore, owing to the different local training or communication time encountered in practical applications, increased attention should be dedicated to asynchronous FL training algorithms for efficiently updating the semantic coding/decoding networks. However, as research progresses, similar to classic Shannonian systems, gradually more and more system parameters should be jointly optimized for finding the entire so-called {\em non-dominated Pareto-front} of all the  optimal system configurations of a two, three- and multi-component objective function.

\subsection{\textbf{Robust Privacy-Preserving Semantic Communication Exhibiting Fairness}}
Multi-user semantic communication is in its infancy at the time of writing, hence it is fragile and vulnerable. The interpretation of the semantic information at the receiver in a multi-user environments is a complex process, relying on joint multi-user detection, channel decoding and semantic decoding. Furthermore, the existing privacy-preserving schemes pay limited attention to robustness and fairness, which must be given cognizance in the design of future semantic communication systems. To address this problem, designs striking an attractive privacy, robustness and fairness trade-off must be found. Nowadays, blockchain is viewed as an underlying platform, which can decentralize the storage, computing and transmission of the Metaverse as well as enhance privacy preservation. In future research, we should attempt to integrate blockchain and semantic communications to realize robust and fair privacy-preserving information interaction by ensuring the authenticity of the semantic data transmitted. Besides, future research has to develop adaptive personalized schemes for heterogeneous scenarios.

\subsection{\textbf{Heterogeneous Privacy-Preserving Semantic Communication}}
The heterogeneity of semantic communications in the Metaverse manifests itself in terms of non-i.i.d data, inconsistent KBs and task-specific encoding/decoding networks, which imposes design challenges. Although these KBs may be rendered more homogeneous by sharing them among clients, this would be a resource-intensive high-latency process. Furthermore, the privacy of the shared KBs will be inevitably eroded. However, most transmitted messages are not dominated by sensitive information. It should be critically appraised, whether the local heterogeneous KBs could be partitioned into public and private segments, so that the clients can upload most messages without any privacy concerns and only the private information would be processed further, as we proposed in Fig. \ref{FLSC}. Indeed, the design of heterogeneous privacy preserving semantic communication schemes for the Metaverse is a wide open issue.

\section{Summary and Conclusions}
We commenced with an overview of the Metaverse assisted by semantic communications. To support trust-worthy semantic communications in the Metaverse, we presented a secure multi-user semantic communication system based on IR as well as GAN and pointed out its drawbacks. Furthermore, we constructed a FL-aided multi-user semantic communication system for the Metaverse, which strikes a balance between privacy and accuracy. Since the FedAvg is unsuitable for heterogeneous KBs and neural networks, we adopt FD instead of FedAvg to obtain all aggregated encoder/decoder model. By harnessing a lightweight DP, the privacy protection is further enhanced. Additionally, the open research issues are summarized to highlight the challenges both in long-term theoretical research and in practical implementations of trust-worthy semantic communications in the Metaverse. Finally, we highlighted the privacy preservation issues in semantic communication applications for the Metaverse, to inspire pioneering research in this emerging area.

\bibliographystyle{IEEEtran}
\bibliography{ref-WCM}

\begin{thebibliography}{10}
\providecommand{\url}[1]{#1}
\csname url@samestyle\endcsname
\providecommand{\newblock}{\relax}
\providecommand{\bibinfo}[2]{#2}
\providecommand{\BIBentrySTDinterwordspacing}{\spaceskip=0pt\relax}
\providecommand{\BIBentryALTinterwordstretchfactor}{4}
\providecommand{\BIBentryALTinterwordspacing}{\spaceskip=\fontdimen2\font plus
\BIBentryALTinterwordstretchfactor\fontdimen3\font minus
  \fontdimen4\font\relax}
\providecommand{\BIBforeignlanguage}[2]{{%
\expandafter\ifx\csname l@#1\endcsname\relax
\typeout{** WARNING: IEEEtran.bst: No hyphenation pattern has been}%
\typeout{** loaded for the language `#1'. Using the pattern for}%
\typeout{** the default language instead.}%
\else
\language=\csname l@#1\endcsname
\fi
#2}}
\providecommand{\BIBdecl}{\relax}
\BIBdecl

\bibitem{she2023}
Z.~Meng, C.~She, G.~Zhao, and D.~De~Martini, ``Sampling, communication, and
  prediction co-design for synchronizing the real-world device and digital
  model in metaverse,'' \emph{IEEE Journal on Selected Areas in
  Communications}, vol.~41, no.~1, pp. 288--300, Jan. 2023.

\bibitem{wang2023}
Y.~Wang, Z.~Su, N.~Zhang, R.~Xing, D.~Liu, T.~H. Luan, and X.~Shen, ``A survey
  on metaverse: Fundamentals, security, and privacy,'' \emph{IEEE
  Communications Surveys \& Tutorials}, vol.~25, no.~1, pp. 319--352, 2023.

\bibitem{blockchain}
Y.~Lin, Z.~Gao, H.~Du, D.~Niyato, J.~Kang, R.~Deng, and X.~S. Shen, ``A unified
  blockchain-semantic framework for wireless edge intelligence enabled web
  3.0,'' \emph{IEEE Wireless Communications, (DOI: 10.1109/MWC.018.2200568)},
  2023.

\bibitem{semantic1}
X.~Luo, H.-H. Chen, and Q.~Guo, ``Semantic communications: Overview, open
  issues, and future research directions,'' \emph{IEEE Wireless
  Communications}, vol.~29, no.~1, pp. 210--219, Feb. 2022.

\bibitem{SCxie}
H.~Xie, Z.~Qin, G.~Y. Li, and B.-H. Juang, ``Deep learning enabled semantic
  communication systems,'' \emph{IEEE Transactions on Signal Processing},
  vol.~69, pp. 2663--2675, Apr. 2021.

\bibitem{attention}
Y.~Wang, M.~Chen, T.~Luo, W.~Saad, D.~Niyato, H.~V. Poor, and S.~Cui,
  ``Performance optimization for semantic communications: An attention-based
  reinforcement learning approach,'' \emph{IEEE Journal on Selected Areas in
  Communications}, vol.~40, no.~9, pp. 2598--2613, Sept. 2022.

\bibitem{kalfa2021towards}
M.~Kalfa, M.~Gok, A.~Atalik, B.~Tegin, T.~M. Duman, and O.~Arikan, ``Towards
  goal-oriented semantic signal processing: Applications and future
  challenges,'' \emph{Digital Signal Processing}, vol. 119, p. 103134, Dec.
  2021.

\bibitem{NLP}
N.~Farsad and A.~Goldsmith, ``Neural network detection of data sequences in
  communication systems,'' \emph{IEEE Transactions on Signal Processing},
  vol.~66, no.~21, pp. 5663--5678, Nov. 2018.

\bibitem{text}
H.~Xie, Z.~Qin, G.~Y. Li, and B.-H. Juang, ``Deep learning enabled semantic
  communication systems,'' \emph{IEEE Transactions on Signal Processing},
  vol.~69, pp. 2663--2675, Apr. 2021.

\bibitem{speech}
Z.~Weng and Z.~Qin, ``Semantic communication systems for speech transmission,''
  \emph{IEEE Journal on Selected Areas in Communications}, vol.~39, no.~8, pp.
  2434--2444, Aug. 2021.

\bibitem{zhang2022deep}
H.~Zhang, S.~Shao, M.~Tao, X.~Bi, and K.~B. Letaief, ``Deep learning-enabled
  semantic communication systems with task-unaware transmitter and dynamic
  data,'' \emph{arXiv preprint arXiv:2205.00271}, 2022.

\bibitem{table1}
J.~Srinivas, A.~K. Das, N.~Kumar, and J.~J. P.~C. Rodrigues, ``Cloud centric
  authentication for wearable healthcare monitoring system,'' \emph{IEEE
  Transactions on Dependable and Secure Computing}, vol.~17, no.~5, pp.
  942--956, Sept. 2020.

\bibitem{table2}
S.~A. Moqurrab, A.~Anjum, N.~Tariq, and S.~Gautam, ``Instant-anonymity: A
  lightweight semantic privacy guarantee for 5{G}-enabled {II}o{T},''
  \emph{IEEE Transactions on Industrial Informatics}, vol.~19, no.~1, pp.
  951--959, Jan. 2023.

\bibitem{table3}
X.~Luo, Z.~Chen, M.~Tao, and F.~Yang, ``Encrypted semantic communication using
  adversarial training for privacy preserving,'' \emph{arXiv preprint
  arXiv:2209.09008}, 2022.

\bibitem{distill}
L.~Liu, J.~Zhang, S.~H. Song, and K.~B. Letaief, ``Communication-efficient
  federated distillation with active data sampling,'' in \emph{IEEE
  International Conference on Communications}, Seoul, Korea, Aug. 2022, pp.
  201--206.

\end{thebibliography}

\vspace{-13 mm}

\begin{IEEEbiographynophoto}{Jianrui Chen} received his B.S. degree in electronics and information engineering from the Jilin University, Changchun, China in 2019, and the M.S. degree in electronic and information engineering from Tsinghua University, Beijing, China in 2023. He is currently pursuing the Ph.D. degree in the School of Cyber Science and Technology from Beihang University, Beijing, China. His research interests lie in the area of wireless communications and security.
\end{IEEEbiographynophoto}

\vspace{-12 mm}

\begin{IEEEbiographynophoto}{Jingjing Wang (corresponding author)} received his B.S. degree in electronic information engineering from Dalian University of Technology, Liaoning, China, in 2014 and Ph.D. degree in information and communication engineering from Tsinghua University, Beijing, China in 2019, both with the highest honors. He is an professor at the School of Cyber Science and Technology, Beihang University, Beijing, 100191, China. His research interests include artificial intelligence-enhanced next-generation wireless networks, swarm intelligence, and confrontation.
\end{IEEEbiographynophoto}

\vspace{-12 mm}

\begin{IEEEbiographynophoto}{Chunxiao Jiang} is an associate professor in School of Information Science and Technology, Tsinghua University. He received the B.S. degree in information engineering from Beihang University, Beijing in 2008 and the Ph.D. degree in electronic engineering from Tsinghua University, Beijing in 2013, both with the highest honors. His research interests include application of game theory, optimization, and statistical theories to communication, networking, and resource allocation problems, in particular space networks and heterogeneous networks.
\end{IEEEbiographynophoto}

\vspace{-12 mm}
\begin{IEEEbiographynophoto}{Yong Ren} received his B.S, M.S and Ph.D. degrees in electronic engineering from Harbin Institute of Technology, China, in 1984, 1987, and 1994, respectively. Now he is a full professor of Department of Electronic Engineering and serves as the director of the Complexity Engineered Systems Lab in Tsinghua University. His current research interests include marine information network, swarm intelligence and wireless AI.
\end{IEEEbiographynophoto}

\vspace{-12 mm}

\begin{IEEEbiographynophoto}{Lajos Hanzo} (\url{http://www-mobile.ecs.soton.ac.uk}, \url{https://en.wikipedia.org/wiki/Lajos_Hanzo}) (FIEEE'04) received Honorary Doctorates from the Technical University of Budapest and Edinburgh University. He is a Foreign Member of the Hungarian Science-Academy, Fellow of the Royal Academy of Engineering (FREng), of the IET, of EURASIP and holds the IEEE Eric Sumner Technical Field Award.
  \end{IEEEbiographynophoto}

\end{document}